\documentclass[rmp,twocolumn]{revtex4}

\usepackage{graphicx,txfonts}

\begin{document}

\title{Efficiency of navigation in indexed networks}

\author{Petter Holme}
\affiliation{Department of Computer Science, University of New Mexico,
  Albuquerque, NM 87131, USA}

\begin{abstract}
  We investigate efficient methods for packets to navigate in complex
  networks. The packets are assumed to have memory, but no previous
  knowledge of the graph. We assume the graph to be indexed,
  i.e.\ every vertex is associated with a number (accessible to the
  packets) between one and the size of the graph. We test different
  schemes to assign indices and utilize them  in
  packet navigation. Four different network models with very different
  topological characteristics are used for
  testing the schemes. We find that one scheme outperform the others,
  and has an efficiency close to the theoretical optimum. We discuss the use of
  indexed-graph navigation in peer-to-peer networking and other
  distributed information systems.
\end{abstract}

\maketitle

\section{Introduction}

The interplay between network structure and search
dynamics has emerged as a busy sub-field of statistical network studies
(see e.g.\ Refs.~\cite{klei:nav1,ada:se2,bjk:pfs,sen:sea,zhu:sea}). Consider a simple
graph $G=(V,E)$ (where $V$ is a set of $n$ vertices and $E$
is a set of $m$ edges---unordered pairs of vertices). Assume information
packets travel from a source vertex $s$ to a destination $t$. We assume
the packages are myopic agents (at a given timestep they have access to
information about the vertices in their neighborhood, but not more),
have memory (so they can e.g.\ perform a depth-first search) but no
previous knowledge of the network. Let $\tau(p)$ be the
time for a packet $p$ to travel between its source and destination. One
commonly studied quantity of search efficiency is the expectation value of $\tau$,
$\bar\tau$, for randomly chosen $s$ and $t$. In this work we attempt to find efficient ways to index $V$ and utilize these indices for packet navigation.

We propose two schemes of indexing the vertices, and
corresponding methods for packet navigation. These schemes, along with two
depth-first search methods (not using vertex indices for more than remembering the path) are examined on
four network models. We will first present the indexing and search
schemes, then the network models for testing the algorithms, and last numerical results.

\begin{figure}
  \centering\includegraphics[width=\linewidth]{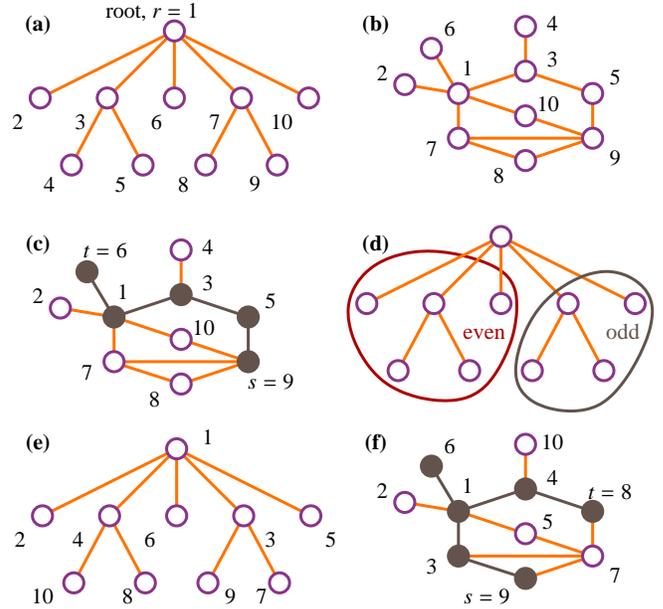}
  \caption{Illustration of the ASD (panels (a)--(c)) and ASU (panels
    (d)--(f)) indexing and search schemes. (a) shows a search tree
    where a local search algorithm can find the shortest path from one vertex to another fast. (b) shows a network indexed by the ASD
    scheme. The tree used in the construction is identical to the one
    shown in (a). Panel (c) shows an ASD search from $s$ to $t$ (with
    $\tau=4$). On the way from $t$ to $r$ the packet chooses the
    neighbor (of the current vertex) with lowest index, which here
    gives a longer route than the optimal $\{(9,10),(10,1)\}$. (d)
    shows a possible partition of branches of non-root vertices into
    classes of as similar size as possible (as done in the ASU indexing
    scheme). (e) shows a possible indexing based on the partition in
    (d). Panel (f) displays a search from $s$ to $t$ with
    $\tau=6$. The shortest path from $t$ to $r$ is accurately found,
    but a detour to $6$ makes the search from $r$ to $t$ sub-optimal.
  }
\label{fig:ill}
\end{figure}

\section{Indexing and search schemes}

Now we turn to the schemes for assigning indices to the vertices and
using them in search processes. Our two main schemes are both inspired
by search trees. Packets first 
moves towards a root vertex $r$, then towards the destination. Unless
the network really is a tree, this approach cannot be exact---a packet
is not guaranteed to find the shortest way both from $s$ 
to $r$ and from $r$ to $t$). However, as we will see, one can assign indices
such that the search either from $s$ to $r$, or from $r$ to $t$ is
certain to be as short as possible. One of our schemes, ASD (accurate
search up),  will be such that the shortest upward search is
guaranteed, the other, ASD (accurate search down), will have the
shortest possible $r$ to $t$ search.

On a technical note, $V$ is
a set of distinct elements and an indexing scheme is a
bijection $\phi:V\mapsto [1,n]$. In the remainder of the text we
will not explicitly distinguish $i\in V$ from $\phi(i)$.

\subsection{The ASD indexing and search}

The numbers $1,\cdots,n$ can be arranged into a search
tree~\cite{algorithms} such that the expected value of $\tau$ scales
like $\log n$. In Fig.~\ref{fig:ill}(a) we give an example of a search
tree. To go from source $s$ to destination $t$ a packet first moves to
the root $r$ by going to the neighbor with lowest index value. From the
root to the destination, the package moves to the neighbor with the
largest index smaller than, or equal to, $t$. Our strategy for the
ASD indexing and search scheme is to construct a spanning
tree $T(G)$ for the network; index the tree to make it a search tree; and use
the algorithm above to navigate from $s$ to $t$. The problem is,
however that real networks are not trees. Imagine adding edges between
vertices of the same heights and branches to the tree in
Fig.~\ref{fig:ill}(a)---the tree will still be a spanning tree, but
the packets may not take the same path from $s$ to $t$ any
more. As we will see, with certain ways of constructing the tree and
indexing the vertices the search, either from $s$ to $r$ or $r$ to $t$
will be optimal.

We construct $T(G)$ in the following way:
\begin{enumerate}
\item Let the root $r$ be a vertex of smallest eccentricity (maximal
  distance to an other vertex).
\item Construct the tree such that the distances to the root is the
  same in $T(G)$ and $G$. In other words, such that all edges in $T$ go between
  different neighborhoods $\Gamma_l(r)=\{i\in V: d(i,r)=l$ and
  $\Gamma_{l+1}(r)\}$ for some level $0\leq l\leq h$, where $h$ is the
  \textit{height} of the tree (by the choice of $r$, $h$ is also the
  radius of the graph). Such a tree can be constructed by finding the set of followed edges in a breadth-first
  search~\cite{algorithms} starting from $r$.
\end{enumerate}
When it is not clear which vertex, or edge, to choose in the above
construction, we choose one at random from all the possible candidates.
When $T$ is constructed, let the indices be a preordering of the vertices in $T$ (i.e.\ the
order of first-occurrence of the vertex in a depth-first search of the
graph)~\cite{algorithms}.

Now we prove that this indexing and search algorithm always gives the
shortest paths from the root to a vertex $t$.
Let $E_T$ be the edges of $T$ and let $T_i$ be the maximal
subtree with $i$ as root. By construction, all vertices in $T_i$ have
indices in $[i,i+|T_i|]$ (where $|\;\cdot\;|$ denotes the cardinality
of a subgraph). Let $i'$ be the largest index in $i$'s neighborhood
smaller than $t$. Assume there is an edge $(i,j)\in E\setminus E_T$
that the search will follow, i.e.\ that $i'<j<t$. This means that
$j\in T_{i'}$. By construction, $i'$ is the only vertex in $T_{i'}$ at
a distance $d(r,i')$ (the distance from the rest of $T_{i'}$ to the root is at least
$d(r,i')+1$). Since $d(r,i') = d(r,i)+1$, we have $d(r,j)\geq d(r,i)+2$
which contradicts the existence of an edge $(i,j)\in E$. Thus searches
from $r$ to $t$ will always follow the edges of $T$, which also means
the $r$--$t$-searches will be as short as possible.

Searching upwards, from $i$ to $r$, in a graph indexed as above is
harder. We know that one shortest path goes via a vertex $j$ with
smaller index than $i$, but there might sub-optimal paths via 
indices $i'$ in the intervals $r<i'<j$ and $j<i'<i$, and there
might also be paths via vertices of index larger than $j$, that is optimal.
For example, assume the search tree in Fig.~\ref{fig:ill}(a)
comes from a graph with the additional edges $(5,9)$, $(8,9)$ and
$(9,10)$ (see Fig.~\ref{fig:ill}(b)). Then, the shortest path from $9$ to $r$ via a vertex of
lower index is $\{(9,7),(7,1)\}$, but there is an equally long path
via a vertex of larger index, $\{(9,10),(10,1)\}$, and longer paths
via vertices both smaller and larger than $7$ but smaller
than $9$. There thus no general way of finding the shortest way from
$s$ to $r$. Instead, we always choose the vertex with the smallest
index in the neighborhood. By this strategy a packet will come closer to $r$, in
index space, for every step. Furthermore, in tree-like parts of the
graph, the search will follow the shortest paths. An illustration of
the ASD search can be found in Fig.~\ref{fig:ill}(c).

\subsection{The ASU indexing and search}

Consider a tree $T(G)$ constructed as in the previous section and an
indexing such that $d(i,r) < d(j,r)$ implies $i<j$ (i.e., all indices
of a level further from the root is larger than in levels closer to
$r$). With such an indexing, since the neighbor of a vertex with the
smallest index necessarily is one step closer to the root, a packet
can always find one shortest way too the root. But once the package is
at the root the indices is not of so much help. The search from $r$ to
$t$ has to be, essentially, a depth-first search. There are, however,
a few tricks to speed up the search. First, there is no need to
search deeper than $t$---if $j>t$, then $t\notin T_j$. Second, one can
choose the indices $i,\cdots,i+|\Gamma_l(r)|$ of one level in the tree
in a way to narrow down the search. For example, one can divide the
vertices into $\nu$ classes (defined by e.g.\ the remainder when the
index is divided by $\nu$) and index vertices of connected regions of
the graph with indices of the same class. The search can then be
restricted to the same class as the destination. We will pursue this idea
with $\nu=2$.

To derive the ASU indexing scheme, the first goal is to divide the vertices into classes of
connected subgraphs. Furthermore, we require all classes to be connected to the root vertex. Another aim is to make the classes of as similar sizes as
possible. Our first step is to make $k_r$ (the degree, or number of
neighbors, of $r$) parallel depth-first searches\footnote{Every
  iteration, one step is taken in all branches. The different search branches marks the
  visited vertices with their indices. A search proceeds only to
  vertices not marked by any search. When there are no unmarked vertices, the search branch is finished.}. Second we group the
$k_r$ search trees into $\nu$ groups with maximally similar sizes. In our case, we seek a partition of the search trees into
two classes such that the sums of vertices in the respective classes
are as close as possible.\footnote{We do this by randomly exchanging
  search trees between the two classes and accept changes that improve
  the partition. The search is continued until their vertex-sums differ
  by at most one, or the partition has not improved for 1000
  trials.} Then we go through the levels, starting from the root,
and assign numbers such that vertices of one partition have even
indices, while the other has odd numbers (this assignment might not
always work). To avoid systematic errors we sample the elements of
levels randomly. This construction scheme is illustrated in
Fig.~\ref{fig:ill}(d) and (e).

\subsection{Degree-based and random search}

As a reference, we also run simulations for two depth-first search
methods that do not utilize indices~\cite{ada:se2}. One of them,
\textsc{Rnd}, is regular depth-first search where the neighbors are
traversed in random order. In the other, \textsc{Deg}, the neighbors
are chosen in order from high to low degree. Just like for ASU and ASD
methods, a packet is assumed to have knowledge about its neighborhood---if the
destination is in the neighborhood of a vertex, then the search will be
over the next time step.

\section{Network models}

The efficiency of our indexing and search schemes are more or
less directly affected by the network structure. To investigate this relationship
we test the search schemes on four different types of network models:
modified Erd\H{o}s--R\'enyi (ER) graphs~\cite{er:on}, square lattices,
Barab\'asi--Albert (BA)~\cite{ba:model} and Holme--Kim
(HK)~\cite{hk:model} networks. To facilitate comparison, we have the same
average degree, four (dictated by the square grid), in all networks.

\subsection{Modified ER graphs}

The ER model is the simplest model for randomly generating simple
graphs with $n$ vertices and $m$ edges. The edges are added one by one
to randomly chosen vertex pairs (the only restriction being that
loops or multiple edges are not allowed). A problem for our purpose is
that ER graphs are not necessarily connected (something required to measure $\bar\tau$). To remedy this we propose a scheme to
make networks connected.
\begin{enumerate}
\item \label{step:components} Detect the connected components.
\item \label{step:seq} Go through the connected components sequentially. Denote
  the current component $C_I$.
\begin{enumerate}
\item \label{step:random_component} Pick a component $C_J$ randomly.
\item Pick a random edge $(i,j)$ whose removal would not fragment
  $C_J$. If no such edge exist, go to step~\ref{step:seq}.
\item \label{step:rnd_vertex} Pick a random vertex $i'$ of $C_I$.
\item \label{step:repl} Replace $(i,j)$ by $(i',j)$. If the edge
  $(i',j)$ would exist already (an unlikely event), go to
  step~\ref{step:repl}. If there is no vertex $i'\in C_I$ such that
  $(i',j)$ does not already exist, then go to~\ref{step:seq}.
\end{enumerate}
\item If the network is disconnected still, go to step~\ref{step:components}.
\end{enumerate}
In practice, even for our largest system sizes, the above algorithm
converges in a few iterations. The number of edges needed to be added
never exceed a few percent of $m$, and this addition is made with greatest possible
randomness; hence we believe the essential network structure of the ER
model is conserved.

\subsection{Square lattice}

We use square lattices with periodic boundary conditions. $n$ vertices
spread out regularly on a $L\times L$-grid such that the vertex with
coordinates $(x,y)$, $1\leq x,y\leq L$, is connected to $(x,y+1)$,
$(x+1,y)$, $(x,y-1)$, $(x-1,y)$ (if $x=1$, we formally let $x-1=L$, if
$x=L$ we let $x+1$ represent $1$; and correspondingly for $y$).

\subsection{BA model}

The popular BA model~\cite{ba:model} of networks with a power-law degree distribution
works as follows (with our parameter settings). Start with one vertex
connected to two degree-one vertices. Iteratively add vertices
connected to two other vertices. Let the probability of connecting the new vertex to
a vertex $i$ already present in the network is proportional to $k_i$
(so called \textit{preferential attachment}).

\subsection{HK model}

The HK model~\cite{hk:model} is a modification of the BA model to give the network
higher number of triangles. When edges are added from the new vertex
to already present vertices, the first edge is added by preferential attachment. The second edge is added to one of $i$'s neighbors, forming a triangle.

\begin{figure}
  \centering
  \includegraphics[width=0.92\linewidth]{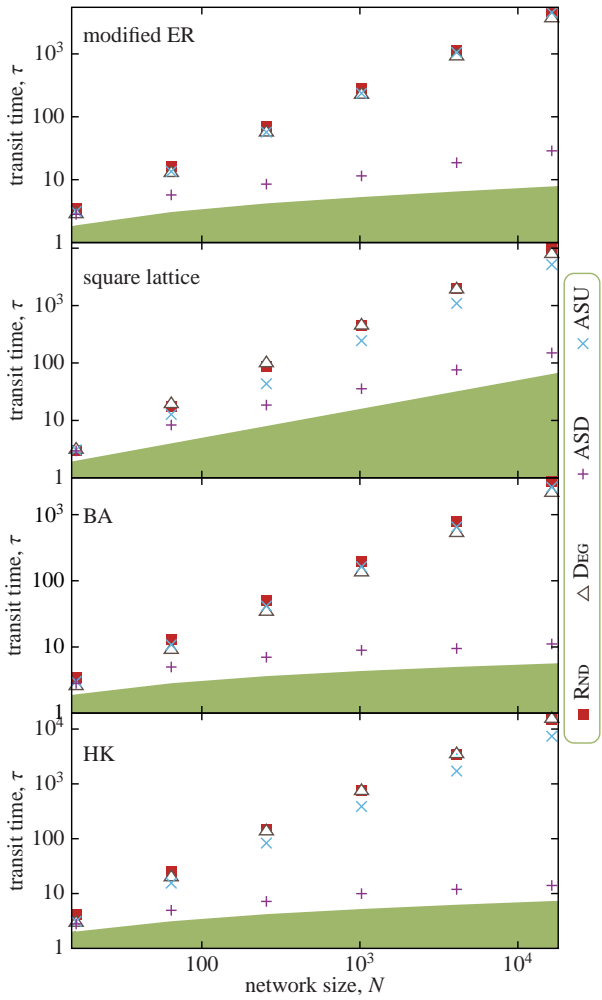}
  \caption{The average search time $\bar\tau$ as a function of the
    graph sizes $n$. In all panels, we display data for the
    different indexing and search schemes. The shaded areas are
    unreachable (corresponding to $\bar\tau$ values smaller than the theoretical minimum---the average distance $\bar{d}$). The different panels correspond to the modified ER
    model, square grid, BA model and HK model networks
    respectively. Error bars would have been smaller than the symbol
    sizes.
  }
\label{fig:sca}
\end{figure}

\section{Numerical results}

We study the search schemes on the four
different network topologies numerically. We use $100$ independent
networks and $100$ different $s$--$t$-pairs for every network. The
network sizes range from $n=16$ to $n=16{,}384$.

\begin{figure}
  \centering
  \includegraphics[width=0.8\linewidth]{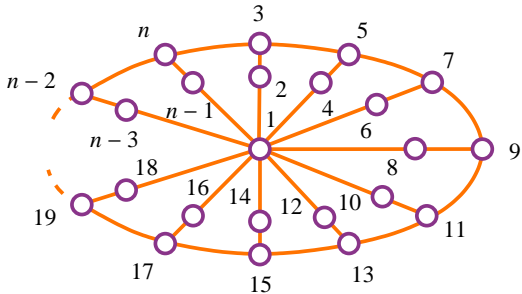}
  \caption{A worst case scenario for navigating from $s$ to $r$ with
    the ASD indexing and search scheme. A packet from $n-2$ to $1$ will
    travel along the perimeter to $3$ and then move towards the
    center.
  }
\label{fig:wcs}
\end{figure}

In Fig.~\ref{fig:sca} we display the average search times
as a function of system size for our simulations. The most conspicuous
feature is that the ASD scheme is always, by far, the most efficient.
While ASU and \textsc{Deg} are close to the least efficient
method (\textsc{Rnd}), ASD is rather close to the theoretical limit
(equal to the average distances $\bar\tau$---the upper border of the shaded
areas in Fig.~\ref{fig:sca}). To be more precise, $\bar\tau$ is quite
constant, about two times larger than the average distance. The other
search schemes (ASU, \textsc{Deg} and \textsc{Rnd}) follow faster
increasing functional forms. For the square lattice, these three schemes
increase, approximately proportional to $n$ (the analytical value
for two-dimensional random walks) whereas for ASD, $\bar\tau$ scale
like distances in square grids, $n^{1/2}$. One way of interpreting this result is that while ASD
manages to find the root as fast as it finds the destination from the root,
ASU fails to find $t$ faster than a random search. The slow
downward performance of ASU is not unexpected---the $r$--$t$-search in
ASU only differs from a random
depth-first search in that it does not search further than the level
of the destination, and that it restricts the search-space to half its original
size by dividing the vertices into odd and even indices. The fast
upward search of ASD is more surprising. In Fig.~\ref{fig:wcs} we show
a network where ASD performs badly. The average time to search upwards
is $(n^2+20n-13)/8n\rightarrow n/8$ as $n\rightarrow\infty$. The
downward search takes $3(n-1)/2n \sim 3/2$ giving a total expected
value of $\bar\tau \sim n/8$. This can be compared to
the average distance $\bar{d}=3-21/4n+2/n^2 \sim 3$. For this example,
$\bar\tau$ and $\bar{d}$ diverge in a way not seen in the network
models. Why is the search so much faster in the model networks? One
point is that the worst-case indexing seen in Fig.~\ref{fig:wcs} is
very unlikely. Since the spokes would be sampled randomly, the chance that a
vertex at the perimeter not finds $r$ in two steps is $1/2$, the
probability of a perimeter vertex to find $r$ in $3$ steps is
$1/4$, and so on. Carrying on this calculation, a vertex at the
perimeter reaches $r$ in $2\sum_k k 2^{k}+2\sim 6$ timesteps giving
$\bar\tau\sim 5$---not too far from the observed $\bar\tau/\bar{d}\sim 2$. We note however that for the model
networks many other factors that are not present in
the wheel-graph of Fig.~\ref{fig:wcs} affect $\bar{\tau}$. For example, the high density
of short triangles in the HK model networks will introduce many edges
between vertices of the same level in $T(G)$ which will affect the
search efficiency.

$\bar\tau$ is approximately linear for the ASU, \textsc{Deg} and
\textsc{Rnd} on all network models. The slopes of these curves are, however, a little different. First, the \textsc{Deg} method is more efficient (compared to
ASU and \textsc{Rnd}) for BA networks, than for the modified ER model. This
observation (also made in Ref.~\cite{ada:se2}) can be explained by the
skewed degree distribution in the BA-network---the packet reaches
high-degree vertices fast. The packet can see a
large part of the network from these hubs, and is therefore more likely to see $t$. More interesting, perhaps, is that ASU is more efficient
for the networks with a higher density of short cycles (the square
lattice and HK models). A rough explanation is that the partition
procedure of ASU cuts off many edges between vertices at the same distance
from $r$. Since there are many such edges in these network models, the network will effectively be sparser
(without changing $G$'s diameter), which results in a better
performance.

\section{Discussion}

We have investigated navigation in valued graphs, more specifically
in indexed graphs---graphs where every vertex is associated with a
unique number in the interval $[1,n]$. These indices can be assigned
to facilitate the packet navigation. The packets are assumed
to have no \textit{a priori} knowledge about the network, except the neighborhoods of their current positions, but memory
enough to perform a depth-first search. We find that one of our
investigated methods, ASD, is very efficient for four topologically
very different network models. The searches with the ASD scheme are
roughly twice as long as the shortest paths (scaling in the same way as the average distance).

Navigation on indexed graphs has applications in distributed
information systems. If, specifically, the amount of information that can be
stored at the vertices were limited, search strategies such as ours
would be useful. One such system is the Autonomous System level Internet where the information stored at each vertex (with the current protocols) increase at least as fast as the networks themselves. For most real-world applications (other examples being \textit{ad hoc}
networks\cite{adhoc} or peer-to-peer networks~\cite{sarshar,niloy,mejngg:p2p}) there are additional
constraints so that the algorithms of this paper cannot immediately be
applied. Such networks are typically changing over time, so the
indexing should ideally be possible to be extended on the fly as vertices
and edges are added and deleted from the network. Apart from this, a
future direction for research on indexed graphs is to improve the
performance of the algorithms presented in this work. There might be
search-tree based algorithm that neither finds the shortest path to
the root, nor finds the shortest way to the destination. For some
network topologies there might be faster algorithms that are not based
on constructing a spanning tree. Consider, for example, modular
networks~\cite{mejn:commu} (i.e.\ networks with tightly connected
subgraphs that are only sparsely interconnected) in such networks the
search can be divided into two stages---first find the cluster of the
destination, then the destination. These two stages should be
reflected in a fast navigation algorithm.

\acknowledgments{
  PH acknowledges financial support from the Wenner-Gren
  Foundations and the National Science Foundation (grant
  CCR--0331580).
}

\end{document}